\documentclass[reprint, aps, prfluids, superscriptaddress, onecolumn, longbibliography, 12pt]{revtex4-2}

\usepackage{csquotes}
\usepackage[T1]{fontenc}
\usepackage[utf8]{inputenc}
\usepackage{amsmath}
\usepackage{amssymb}
\usepackage{enumitem}
\usepackage{graphicx}
\usepackage{refstyle}
\usepackage[
  citecolor=blue,
  colorlinks,
  linkcolor=blue,
  urlcolor=blue,
]{hyperref}
\usepackage[dvipsnames]{xcolor}

\newcommand{\be}{\begin{equation}}
\newcommand{\ee}{\end{equation}} 
\newcommand{\bea}{\begin{eqnarray}}
\newcommand{\eea}{\end{eqnarray}}

\begin{document}


\title{{\color{black}Prandtl number dependence of the small-scale properties in turbulent Rayleigh-B{\'e}nard convection}}



\author{Shashwat Bhattacharya}
\email{shabhatt@iitk.ac.in}
\affiliation{Department of Mechanical Engineering, Indian Institute of Technology, Kanpur 208016, India}
\author{Mahendra K. Verma}
\email{mkv@iitk.ac.in}
\affiliation{Department of Physics, Indian Institute of Technology, Kanpur 208016, India}
\author{{Ravi Samtaney}}
\affiliation{Mechanical Engineering, Division of Physical Science and Engineering, King Abdullah University of Science and Technology, Thuwal 23955, Saudi Arabia}
%
\begin{abstract}

We analyze the Prandtl number (Pr) dependence of spectra and fluxes of kinetic energy, as well as the energy injection rates and dissipation rates, of turbulent thermal convection using numerical data. As expected, for a flow with $ \mathrm{Pr} \lessapprox 1 $, the inertial-range kinetic energy flux  is constant, and the kinetic energy spectrum is  Kolmogorov-like ($ k^{-5/3} $).  More importantly, we show that the amplitudes of the kinetic energy fluxes and spectra and those of structure functions increase with the decrease of Pr, thus exhibiting stronger nonlinearity for flows with small Prandtl numbers. Consistent with these observations, the kinetic energy injection rates and the  dissipation rates too increase with the decrease of Pr. {\color{black}Our results are in agreement with earlier studies that report the Reynolds number to be a decreasing function of Prandtl number in turbulent convection.} On the other hand, the tail of the probability distributions  of the local heat flux  grows  with the increase of Pr, indicating  increased fluctuations in the local heat flux with Pr.

\end{abstract}
\maketitle
\section{Introduction}
\label{sec:Introduction}

Turbulence is a complex phenomenon that remains largely unsolved even today. One of the important results of turbulence is due to \citet{Kolmogorov:DANS1941Dissipation,Kolmogorov:DANS1941Structure} that explains the statistics of three dimensional (3D) turbulence under the assumption of homogeneity and isotropy. In Kolmogorov's model, the kinetic energy  injected at large scales  cascades down to intermediate scales, called the inertial range, and then to dissipative scales. Kolmogorov's theory assumes that in the inertial range, no kinetic energy is injected and the viscous dissipation rate is negligible. This leads to the energy cascade rate $\Pi_u$, also referred to as the kinetic energy flux, being constant in the inertial range and equal to the total viscous dissipation rate $\epsilon_u$. Dimensional analysis leads to the following relation for the energy spectrum $E_u(k)$: 
\begin{equation}
E_u(k)= K_\mathrm{KO} (\epsilon_u)^{2/3}k^{-5/3}, 
\label{eq:KolmogorovSpectrum}
\end{equation}
where $k$ is the wavenumber, and $K_{\mathrm{KO}}$ is the Kolmogorov constant~\cite{Frisch:book,Lesieur:book:Turbulence}.

The dynamics become more involved in turbulent thermal convection. In such flows, buoyancy acts at all scales, including the inertial range, and thus can potentially alter the scaling of kinetic energy spectrum and flux, as discussed by \citet{Bolgiano:JGR1959,Obukhov:DANS1959,Procaccia:PRL1989,Lvov:PRL1991,Lvov:PD1992}, and \citet{Kumar:PRE2014}.  Also, refer to reviews~\citet{Ahlers:RMP2009,Lohse:ARFM2010,Chilla:EPJE2012}, and \citet{Verma:book:BDF,Verma:book:ET}  for further details.  A large number of such works focus on an idealized setup called Rayleigh-B{\'e}nard convection (RBC) where a fluid is enclosed between two horizontal walls with the bottom wall kept at a higher temperature than the top wall~\cite{Chandrasekhar:book:Instability}.  RBC is governed by two dimensionless parameters: the Rayleigh number (Ra), which is the ratio of the buoyancy and the dissipative forces, and the Prandtl number (Pr), which is the ratio of kinematic viscosity to thermal diffusivity.

For stably-stratified turbulence, \citet{Bolgiano:JGR1959} and \citet{Obukhov:DANS1959} proposed that the energy and entropy spectra scale as $E_u(k) \sim k^{-11/5}$ and $E_\theta(k) \sim k^{-7/5}$ respectively for $k < k_B$, the Bolgiano wavenumber, where buoyancy forces are strong. The aforementioned scaling will be  henceforth referred to as Bolgiano-Obukhov (BO) scaling. For $k>k_B$, the buoyancy forces are weak and Kolmogorov-Obukhov scalings of $E_u(k) \sim k^{-5/3}$ and $E_\theta(k) \sim k^{-5/3}$ are expected. Using theoretical arguments, \citet{Procaccia:PRL1989}, \citet{Lvov:PRL1991}, \citet{Lvov:PD1992}, and \citet{Rubinstein:NASA1994} proposed the applicability of BO scaling to RBC as well. Researchers have attempted to confirm the above theory with the help of experiments and numerical simulations.  Many researchers have reported BO scaling in RBC~\cite{Kerr:JFM1996,Ashkenazi:PRL1999spectrum,Shang:PRE2001,Wu:PRL1990,Niemela:Nature2000,Chilla:EPL1993Spectra,Zhou:PRL2001} based on their observations, except for convection in the small Pr regime, where Kolmogorov-Obukhov scaling was reported~\cite{Cioni:EPL1995,Takeshita:PRL1996,Camussi:PF1998,Horanyi:IJHMT1999,Frick:EPL2015,Schumacher:PNAS2015,Mishra:PRE2010}.  In a critical review,  \citet{Lohse:ARFM2010} raised doubts on BO phenomenology in RBC.   

Recent works on turbulent RBC show that thermal plumes inject kinetic energy to the system, hence the kinetic energy flux is a non-decreasing function of $ k $, thus ruling out BO scaling for turbulent thermal convection~\cite{Kumar:PRE2014,Verma:NJP2017}. They also observed that in the inertial range of $\mathrm{Pr}=1$ RBC, the kinetic energy injection rates by buoyancy is relatively weak, hence the kinetic energy flux is nearly constant, and the energy spectrum is Kolmogorov-like ($ k^{-5/3} $).   Some other works ~\cite{Mishra:PRE2010,Kaczorowski:JFM2013,Schumacher:PNAS2015,Kunnen:PRE2014,Bhattacharjee:PLA2015}  too reported similar behaviour.  Note however that large-Pr convection exhibits much steeper kinetic energy spectrum due to strong viscous dissipation~\cite{Pandey:PRE2014,Pandey:Pramana2016}.

In this paper, we perform detailed numerical simulations of RBC for a fixed $\mathrm{Ra} = 10^7 $  and $\mathrm{Pr} = 0.02$, $0.1$, $1$, $6.8$, and $100$, and analyze the relative magnitudes of  energy fluxes and spectra. We observe that turbulence gets stronger with the decrease of Prandtl number.  Interestingly, the magnitudes of entropy ($ \theta^2 $, where $ \theta $ is the temperature fluctuation from the conduction profile) spectra and fluxes do not change significantly  with the Prandtl number. Even though we focus on a small set of parameters, the patterns observed here are expected to be valid for a wide range of parameters. Note, however, that there might be a different pattern in the ultimate regime of very large Rayleigh numbers~\cite{Kraichnan:PF1962Convection} on which intense research is still going on.

The structure function is directly related to the energy spectrum~\cite{Ching:book,Verma:book:ET}.  For homogeneous and isotropic 3D turbulence, the third-order velocity structure function scales as $S_3^u(l) \sim -l$, where $l$ is the length scale~\cite{Kolmogorov:DANS1941Dissipation,Kolmogorov:DANS1941Structure}. The higher order structure functions for such flows fit well with the hierarchy model of \citet{She:PRL1994}. Although there have been many studies on structure functions of thermal convection for moderate Prandtl numbers~\cite{Benzi:EPL1994,Benzi:EPL1994b,Ching:PRE2000,Calzavarini:PRE2002,Ching:PRE2008Anomalous,Sun:PRL2006,Kaczorowski:JFM2013,Kunnen:PRE2008,Ching:PRE2013,Lohse:ARFM2010,Ching:PRE2007}, their reported scalings are not very conclusive. Recently, using high resolution numerical simulations of convection for $\mathrm{Pr}=1$, \citet{Bhattacharya:PF2019b} showed that the structure functions of thermal convection scale similarly as that of 3D hydrodynamic turbulence.  Note that these works are for $ \mathrm{Pr} \le 7 $.  In this paper, we analyze the  structure functions for Pr ranging from 0.02 to 100 and show	 that their amplitudes increase with the decrease of Prandtl number, similar to the amplitudes of the energy spectra.

In turbulent convection, buoyancy injects kinetic energy at all scales, including the dissipation range.  {\color{black} \citet{Bhattacharya:PF2019b} showed that for $\mathrm{Pr}=1$ convection, although the modal kinetic energy injection by buoyancy is small at intermediate wavenumbers (as discussed earlier), it adds up to a significant fraction of the total kinetic energy injection when summed over the inertial and dissipative scales.The inertial-range kinetic energy flux is due to the fraction of energy injected only at large scales and hence less than the total kinetic-energy dissipation rate.}  We expect the kinetic-energy dissipation rate and the energy injection rates by buoyancy to {\color{black}exhibit a similar Pr dependence} as the energy flux.  Our numerical simulations are consistent with these arguments.

Even though heat flows from the bottom plate to top plate, there are strong fluctuations in the local heat flux ~\cite{Shang:PRL2003,Kaczorowski:JFM2013,Shishkina:PF2007,Pharasi:PF2016}.  These above works reveal  both negative and positive local  heat fluxes with exponential tails in RBC.  However, the positive heat fluxes dominate the negative ones, leading to a net vertical heat flux.  Similar asymmetry has been observed in many other systems, for example in wave turbulence~\cite{Falcon:PRL2008}. In this paper, we report the Prandtl number dependence of  the probability distribution of the convective heat flux.

In RBC, the {\color{black} global heat transport and the large-scale velocity are quantified by the Nusselt number (Nu) and the Reynolds number (Re) respectively~\cite{Grossmann:PRL2001,Grossmann:JFM2000,Pandey:PF2016}.}  Many experiments and simulations of RBC have been performed to study the variations of these quantities with Pr. These studies revealed that Nu is {\color{black} a very weak function} of Pr for $\mathrm{Pr} \gtrsim 1$~\cite{Verzicco:JFM1999,Ahlers:PRL2001,Silano:JFM2010,Xia:PRL2002} but varies as $\mathrm{Nu} \sim \mathrm{Pr}^{0.14}$ for $\mathrm{Pr} \ll 1$~\cite{Verzicco:JFM1999}. Further, the scaling of Re has been shown to vary from $\mathrm{Re} \sim \mathrm{Pr}^{-0.7}$ for $\mathrm{Pr} \ll 1$ to $\mathrm{Re} \sim \mathrm{Pr}^{-1}$ for $\mathrm{Pr} \gg 1$~\cite{Verzicco:JFM1999,Lam:PRE2002,Silano:JFM2010}. These observations have been explained by several models~\cite{Grossmann:PRL2001,Grossmann:JFM2000,Shraiman:PRA1990,Shishkina:PRF2017}.   In this paper, we will not discuss the scaling of  large-scale quantities. {\color{black}However, it must be noted that our current observations on the variations of the amplitudes of the spectral quantities and the velocity structure functions with Prandtl number are consistent with the aforementioned studies~\cite{Verzicco:JFM1999,Lam:PRE2002,Silano:JFM2010} that report Re to be a decreasing function of Pr.}

The outline of the paper is as follows. In Sec.~\ref{sec:GoverningEquations}, we present the governing equations of RBC and briefly introduce the spectral quantites that will be analyzed in this work. We discuss the simulation details in Sec.~\ref{sec:Numerics}. In Sec.~\ref{sec:Results}, we obtain the Pr dependence on the spectral quantities using our simulation data. In Sec.~\ref{sec:SFs}, we discuss the scaling of velocity structure functions for different Pr. In Sec.~\ref{sec:PDFs}, we study the probability distribution of the convective heat flux. We conclude in Sec.~\ref{sec:Conclusions}.  The entropy spectra and fluxes for various Prandtl numbers are discussed in Appendix A.

\section{Review of energy spectrum, energy flux, and structure functions of RBC}
\label{sec:GoverningEquations}
The governing equations of RBC under the Boussinesq approximation~\cite{Chandrasekhar:book:Instability}  are as follows:
\bea
\frac{\partial \mathbf{u}}{\partial t} + (\mathbf{u} \cdot \nabla) \mathbf{u} & = & -\frac{\nabla \sigma}{\rho_0} + \alpha g \theta \hat{z} + \nu \nabla^2 \mathbf{u}, \label{eq:momentum}\\
\frac{\partial \theta}{\partial t} + (\mathbf{u} \cdot \nabla) \theta & = & \frac{\Delta}{d}u_z + \kappa \nabla^2 \theta, \label{eq:theta}  \\
\nabla \cdot \mathbf{u} & = & 0, \label{eq:continuity} 
\eea  
where $\mathbf{u}$ and $\sigma$ are the velocity and the pressure fields respectively, $\theta$ is the fluctuation of temperature from conduction state, and $\Delta$ and $d$ are the temperature difference and distance respectively between the top and bottom walls.

{\color{black}We nondimensionalize Eqs.~(\ref{eq:momentum}) to (\ref{eq:continuity}) using $d$ as the length scale, $\sqrt{\alpha g \Delta d}$ as the velocity scale, and $\Delta$ as the temperature scale.
The nondimensionalized variables are as follows~\cite{Verzicco:JFM1999,Emran:JFM2008}.
\begin{equation}
\mathbf{u}' = \frac{\mathbf{u}}{\sqrt{\alpha g \Delta d}}, \quad \nabla ' = \nabla d, \quad \theta' = \frac{\theta}{\Delta}, \quad t' = \frac{\sqrt{\alpha g \Delta d}}{d} t, \quad \sigma' = \frac{\sigma}{\rho_o \alpha g \Delta d}.
\label{eq:NonDimensionalization}
\end{equation} 
The governing equations in terms of the nondimensional variables become 
\bea
\frac{\partial \mathbf{u}'}{\partial t'} + \mathbf{u}' \cdot \nabla' \mathbf{u}' &=& -\nabla' \sigma' + \theta' \hat{z} +  \sqrt{\mathrm{\frac{Pr}{Ra}}}\nabla'^2 \mathbf{u}', \label{eq:NDMomentum} \\
\frac{\partial \theta'}{\partial t'} + \mathbf{u}'\cdot \nabla' \theta' &=&  u_z' + \frac{1}{\sqrt{\mathrm{Ra} \mathrm{Pr}}}\nabla'^2 \theta', \label{eq:NDTheta} \\
\nabla' \cdot \mathbf{u}' &=& 0, \label{eq:NDContinuity}
\eea 
}where $\mathrm{Ra}=\alpha g \Delta d^3/(\nu \kappa)$ is the Rayleigh number, and $\mathrm{Pr}=\nu/\kappa$ is the Prandtl number. The Rayleigh and Prandtl numbers are the governing parameters of RBC. {\color{black} The primes for the nondimensional variables will henceforth be dropped for the sake of brevity}.

We now describe the important quantities that will be used for analyzing the statistics of RBC in this paper. These are the kinetic energy spectrum, kinetic energy flux, buoyant energy injection spectrum, viscous dissipation spectrum, and the velocity structure functions.   The kinetic energy spectrum~\cite{Frisch:book,Lesieur:book:Turbulence} is the kinetic energy contained in a wavenumber shell of radius $k$. It is defined as
\begin{equation}
E_u(k) = \frac{1}{2} \sum_{k\leq |\mathbf{k}'| < k+1} |\mathbf{u}(\mathbf{k}')|^2,  \label{eq:KEspectrum}
\end{equation}
where $\mathbf{u}(\mathbf{k})$ is the Fourier transform of the velocity field. The kinetic energy is injected by buoyancy predominantly at large scales, but some energy is also injected at small scales~\cite{Kumar:PRE2015,Verma:NJP2017,Verma:book:BDF,Verma:book:ET,Lohse:ARFM2010}. The energy injection spectrum $\mathcal{F}_B(k)$ {\color{black}in its nondimensional form} is given by~\cite{Lesieur:book:Turbulence,Frisch:book}
\begin{equation}
\mathcal{F}_B(k) = \sum_{|\mathbf{k}|=k} \Re \{\mathbf{u}(\mathbf{k}) \cdot \mathbf{f}^*(\mathbf{k})\} =  \sum_{|\mathbf{k}|=k} \Re \{u_z(\mathbf{k}) \theta^*(\mathbf{k})\}.
\label{eq:Buoyancy_forcing} 
\end{equation}
where $\mathbf{f}=\theta \hat{z}$ is the buoyancy term in the nondimensional momentum equation and * represents the 
complex conjugate. The injected kinetic energy cascades to smaller scales by nonlinear interactions between different velocity modes. This transfer of energy is quantified by energy flux $\Pi_u$, which is the kinetic energy leaving a wavenumber sphere of radius $k_0$. The flux is computed as follows~\cite{Kraichnan:JFM1959,Dar:PD2001,Verma:PR2004}:
\begin{equation}
\Pi_u(k_0) = \sum_{k \geq k_0} \sum_{p<k_0} \delta_\mathbf{k,p+q} \Im(\mathbf{[k \cdot u(q)][u^*(k) \cdot u(p)]}).
\label{eq:flux_MtoM} 
\end{equation}
The kinetic energy gets dissipated predominantly at small scales due to viscosity. This phenomenon is quantified by the viscous dissipation spectrum, which in nondimensional form is given by~\cite{Frisch:book,Lesieur:book:Turbulence}
\begin{equation}
D(k) = 2 \sqrt{\mathrm{\frac{Pr}{Ra}}} k^2 E_u(k),
\label{eq:viscous_Dissipation}
\end{equation}
with the total viscous dissipation rate being $\epsilon_u = \sum_0^\infty D(k)$. In real space, the total viscous dissipation rate is also given by 
\begin{equation}
\epsilon_u = 2 \sqrt{\mathrm{Pr/Ra}} \langle S_{ij} S_{ij} \rangle,
\label{eq:viscous_dissipation_real}
\end{equation}
where $S_{ij}$ is the strain rate tensor~\cite{Landau:book:Fluid}. 
The aforementioned spectral quantites are related to each other by the variable energy flux equation~\cite{Frisch:book,Lesieur:book:Turbulence,Verma:book:ET}, which for a steady state is
\begin{equation}
\frac{d}{dk}\Pi_u(k) = {\mathcal{F}}_B(k)-{D}(k).
\label{eq:VEF_RBC}
\end{equation}
Note that for a steady state, the total kinetic energy injection rate $\sum_0^\infty \mathcal{F}_B(k)$ equals the total viscous dissipation rate $\epsilon_u$~\cite{Frisch:book,Lesieur:book:Turbulence}.

Apart from the energy spectrum, the structure function is another important diagnostics tool to describe turbulence. The velocity structure function of order $q$ is defined as~\cite{Ching:book,Frisch:book,Lesieur:book:Turbulence}
\begin{equation}
S_q^u(l) = \langle [\{\mathbf{u(r+l)-u(r)}\}\cdot \mathbf{\hat{l}}]^q \rangle, 
\label{eq:SF_q}
\end{equation}
where $\mathbf{r}$ and $\mathbf{l}$  are position vectors, and $l=|\mathbf{l}|$. The second-order  velocity structure function is  related to the energy spectrum~\cite{Lesieur:book:Turbulence,Ching:book}. 

In this paper, we compute and compare the above quantities for different Prandtl numbers using data from numerical simulations. In the next section, we discuss the numerical methods employed in our study.

\section{Details of our numerical simulations}
\label{sec:Numerics}

We numerically solve  Eqs.~(\ref{eq:NDMomentum})-(\ref{eq:NDContinuity})   for Pr  from 0.02 to 100 for a  fixed Rayleigh number of $\mathrm{Ra}=10^7$  to study  Pr dependence of turbulent thermal convection.  The simulations were performed on a cubical domain of unit dimension using the finite difference solver SARAS~\cite{Verma:SNC2020,Samuel:JOSS2020}. The solver uses a second-order spatial discretization scheme and employs multigrid method for solving the pressure-Poisson equations.  No-slip boundary conditions were imposed on all the walls, adiabatic boundary conditions on the sidewalls and isothermal boundary conditions on the top and bottom walls. A second-order Crank-Nicholson scheme was used for time-advancement, and the maximum Courant number was kept at {\color{black}0.7}. The maximum time for simulations range from 3 to 101 free-fall time ($t_\mathrm{ND}$) after attaining a steady state.

The grid resolutions were varied from $257^3$ for $\mathrm{Pr}=100$ to $1025^3$ for $\mathrm{Pr}=0.02$. The above grid resolutions ensure that the grid-spacing $\Delta x$ is smaller than the Kolmogorov length scale $\eta=(\nu^3 \epsilon_u^{-1})^{1/4}$ for $\mathrm{Pr} \leq 1$ and the Batchelor length scale $\eta_\theta=(\nu \kappa^2 \epsilon_u^{-1})^{1/4}$ for $\mathrm{Pr}>1$, indicating that the smallest scales of the simulations are adequately resolved. Further, we have a minimum of 7 points in the viscous and thermal boundary layers, satisfying the resolution criterion of \citet{Grotzbach:JCP1983} and \citet{Verzicco:JFM2003}. We validate our simulations by computing the Nusselt number (see Sec.~\ref{sec:PDFs}) for all our runs and ensuring that they are consistent with earlier results~\cite{Pandey:PRE2016,Pandey:PF2016,Bhattacharya:PF2018,Bhattacharya:PF2019,Scheel:PRF2017,Vishnu:PRF2020}. {\color{black} We also compute the Nusselt number averaged over the last half of the free-fall time range for every run and observe it to match with the Nusselt number averaged over the entire free-fall time range, with the deviation being within approximately 2\%. Thus, we verify that our simulations are statistically converged.} The simulation details are summarized in Table~\ref{table:SimDetails}.

\begin{table*}
	\caption{Details of our data obtained {\color{black} from} direct numerical simulations of RBC performed in a cubical box for $\mathrm{Ra} = 10^7$: the Prandtl number (Pr), the grid size, the ratio of the Kolmogorov length scale (for $\mathrm{Pr} \leq 1$) or the Batchelor length scale (for $\mathrm{Pr} > 1$) to the mesh width ($\eta/\Delta x$), the number of grid points in viscous and thermal boundary layers ($N_{\mathrm{VBL}}$ and $N_{\mathrm{TBL}}$ respectively), the Nusselt number (Nu), the number of non-dimensional time units ($t_\mathrm{ND}$), and snapshots over which the quantities are averaged.}
	\begin{ruledtabular}
		\begin{tabular}{c c c c c c c c}
			$\mathrm{Pr}$ &  Grid size & $\eta/\Delta x$ & $N_{\mathrm{VBL}}$ & $N_{\mathrm{TBL}}$ & Nu & $t_{\mathrm{ND}}$ & Snapshots\\
			\hline 
			$0.02$ & $\mathrm{1025^3}$ & 1.45 & $7$ & $48$ & {\color{black}$10.6 \pm 0.6$} & {\color{black}8} & {\color{black}81}\\ 
			$0.1$ & $\mathrm{513^3}$ & 1.52 & $6$ & $20$  & $13.9 \pm 1.1$  & 33 & 66\\
			$1$ &  $\mathrm{257^3}$ & 2.31 & $5$ & $9$  & $16.3 \pm 1.3$ & 101 & 101 \\
			6.8 &  $257^3$ & 2.33 & 6 & 9 & $15.9 \pm 1.2$ & 101 & 101\\
			100 &  $257^3$ & 2.30 & 7 & 9 & $16.8 \pm 1.3$ & 101 & 101
		\end{tabular}
		\label{table:SimDetails}
	\end{ruledtabular}
\end{table*}

We employ the pseudo-spectral code TARANG~\cite{Chatterjee:JPDC2018,Verma:Pramana2013tarang} to compute the spectra and fluxes of kinetic energy. {\color{black}To compute the Fourier transform of velocity fields, we approximate the velocity field using free-slip boundary conditions~\cite{Verma:book:BDF}, and employ Fourier (sine and cosine) expansion of the field. Note that the viscous boundary layer thickness is very small (less than 5\% of the domain size~\cite{Bhattacharya:PF2020}), and the small-scale structures in the boundary layers do not affect the inertial-range properties~\cite{Verma:book:BDF,Verma:book:ET}. Hence, even for flows bounded by rigid walls, the inertial-range spectra and fluxes can be computed using Fourier expansion with reasonable accuracy. Since the boundary layers are thin, the system can be considered homogeneous.
The Fourier expansion of the velocity field is as follows~\citep{Verma:book:BDF}.
\begin{eqnarray}
u_x(x,y,z) &=& \sum_{k_x,k_y,k_z} u_x (k_x,k_y,k_z) 8 \sin (k_xx) \cos (k_yy) \cos (k_zz),
\label{eq:SSS_ux} \\
u_y(x,y,z) &=& \sum_{k_x,k_y,k_z} u_y (k_x,k_y,k_z) 8 \cos (k_xx) \sin (k_yy) \cos (k_zz),
\label{eq:SSS_uy} \\
u_z(x,y,z) &=& \sum_{k_x,k_y,k_z} u_z (k_x,k_y,k_z) 8 \cos (k_xx) \cos (k_yy) \sin (k_zz),
\label{eq:SSS_uz}
\end{eqnarray}
where $k_x= l\pi$, $k_y= m\pi$, and $k_z= n\pi$; $l$, $m$, and $n$ are positive integers. Using the values of $u_x(k_x,k_y,k_z)$, $u_y(k_x,k_y,k_z)$, and $u_z(k_x,k_y,k_z)$, we compute the kinetic energy spectrum and flux using Eqs.~(\ref{eq:KEspectrum}) and (\ref{eq:flux_MtoM}) respectively.}
The total kinetic energy injection and dissipation rates are computed using Eq.~(\ref{eq:viscous_dissipation_real}).

{\color{black}In RBC, since buoyancy is only along the vertical direction, the flow is expected to be anisotropic. Interestingly, however, \citet{Nath:PRF2016} showed via detailed numerical simulations that RBC is nearly isotropic. In the aforementioned work, \citet{Nath:PRF2016} computed the modal kinetic energy as a function of the polar angle $\Theta$ (angle between the buoyancy direction and the wavenumber) and found it to be nearly independent of $\Theta$.  Based on the results of the above work, the minor directional dependence of the modal kinetic energy can be neglected, and we analyze the spectral quantities as functions of the wavenumber shell radius ($k$) in this paper.} 

We use the parallel code \texttt{fastSF}~\cite{Sadhukhan:JOSS2020} to compute the velocity structure functions. {\color{black} Since RBC is nearly homogeneous and isotropic as described earlier, we average the structure functions over the entire domain}. For our computations, we coarse-grain our data to a $256^3$ grid in order to save computational resources. Note that the aforementioned coarse-graining filters the dissipation scale and it does not impact the inertial range scaling. All the spectral quantities and structure functions are averaged over 30 to 100 snapshots taken at equal time intervals after attaining a steady state.

 In the next three sections, we present our numerical results. 

\section{Variation of spectral quantities with Prandtl number}
\label{sec:Results}
In this section, we analyse the Pr dependence of the kinetic energy spectra, kinetic energy fluxes, energy injection rates, and dissipation rates using our numerical data.

\subsection{Kinetic energy spectra and fluxes}
\label{subsec:KESpectrum}
\begin{figure}[t]
	\includegraphics[scale=0.5]{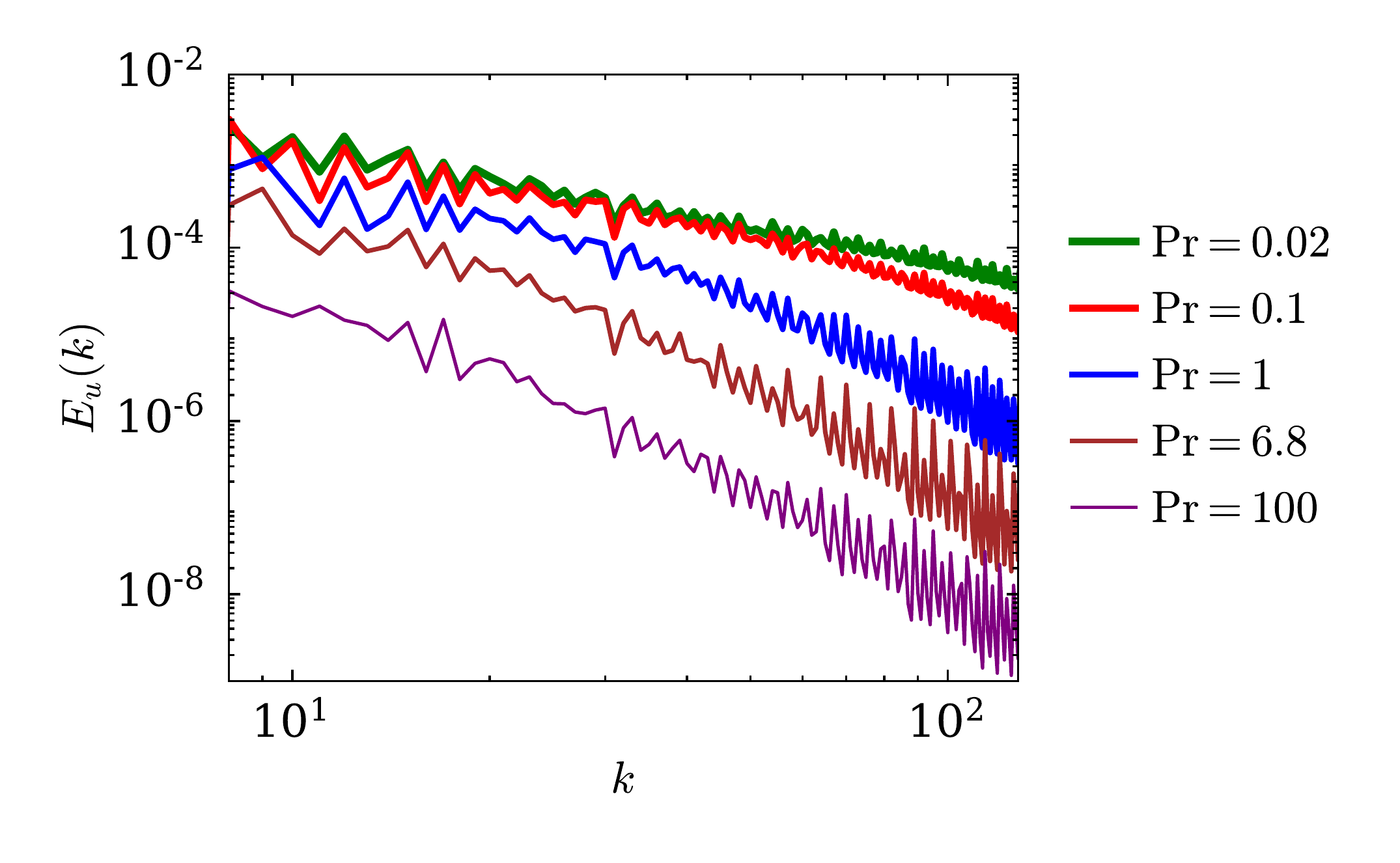}
	\caption{{\color{black}(color online) (a) For $\mathrm{Ra}=10^7$ and $\mathrm{Pr}=0.02$ (green), $0.1$ (red), $1$ (black), $6.8$ (brown), and $100$ (purple): plots of the kinetic energy spectra versus the wavenumber $k$. The spectra exhibits fluctuations at small and intermediate wavenumbers.}} 
	\label{fig:KESpectra}
\end{figure}

We compute the kinetic energy spectrum ($E_u(k)$) and flux ($\Pi_u(k)$) for all our runs using Eqs.~(\ref{eq:KEspectrum},\ref{eq:flux_MtoM}). {\color{black}In Fig.~\ref{fig:KESpectra} exhibits the plots of the kinetic energy spectra for different Prandtl numbers.}
We note that energy spectrum  exhibits fluctuations at small and intermediate wavenumbers that produce large  errors in the best-fit curves~\cite{Stepanov:PRE2014}. To mitigate these errors, we employ  best-fit curves for the \textit{integral energy spectrum}, which is $\int_k^{\infty} E_u(k') dk' = \sum_k^{\infty} E_u(k')$.  The integration process smoothens the curves significantly leading to a major reduction in fitting errors. 

\begin{figure}[b]
	\includegraphics[scale=0.45]{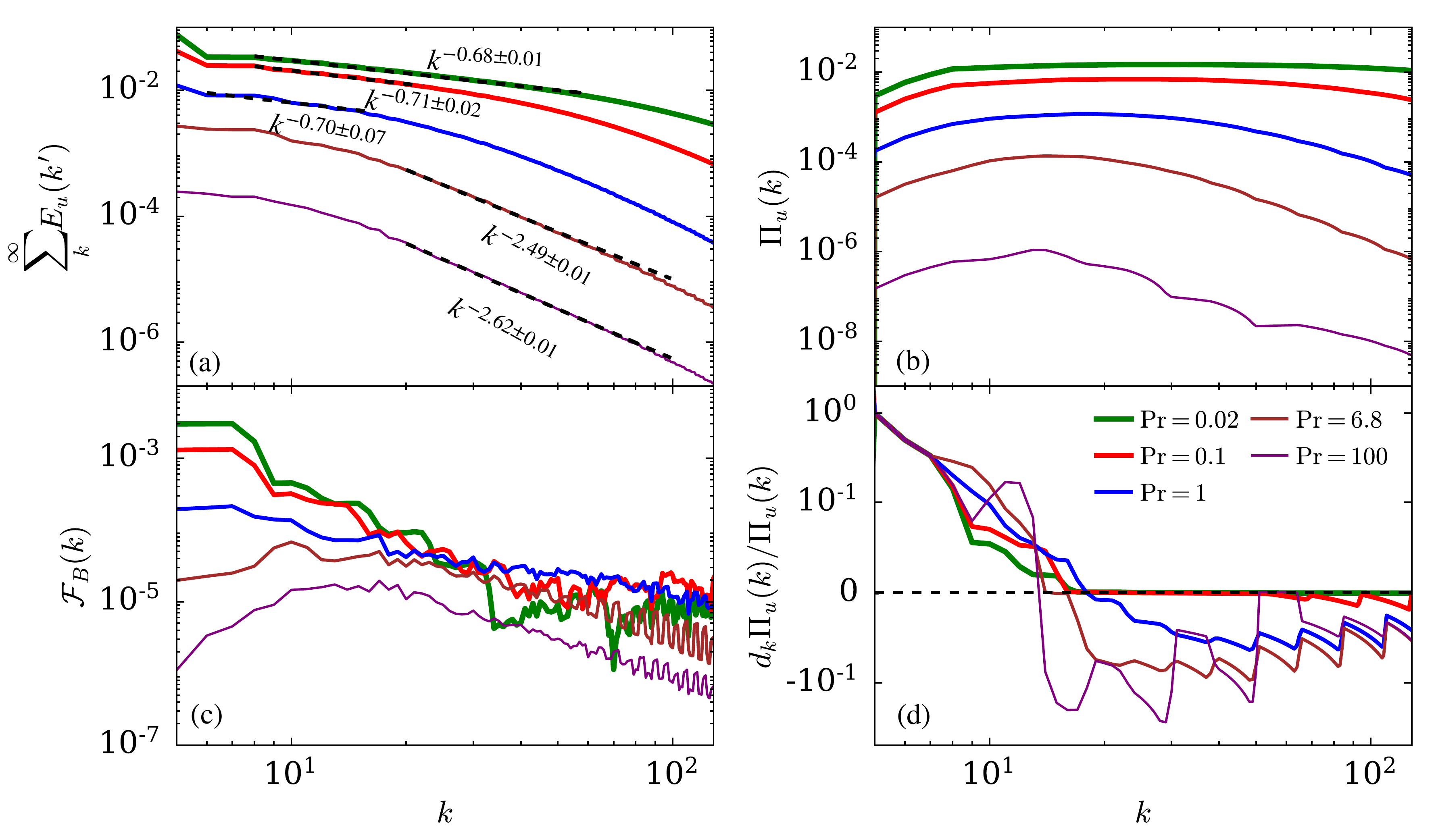}
	\caption{(color online) For $\mathrm{Ra}=10^7$ and $\mathrm{Pr}=0.02$, $0.1$, $1$, $6.8$, and $100$: (a) Integral kinetic energy spectrum, $\sum_k^{\infty} E_u(k')$  vs. wavenumber $k$, (b) kinetic energy flux, $\Pi_u(k)$, vs. $k$, (c) energy injection rate due to buoyancy, $\mathcal{F}_B(k)$, vs. $k$, and (d) $d_k\Pi_u(k)/\Pi_u(k)$ vs. $k$. The amplitudes of the energy spectrum and flux decrease with Pr. For $\mathrm{Pr} \leq 1$, the energy spectrum exhibits Kolmogorov's scaling.} 
	\label{fig:KESpectra_Flux}
\end{figure}

For the inertial-range spectral form of $E_u(k) \sim k^\alpha$, the integral  $\sum_k^{\infty} E_u(k') \sim k^{\alpha+1}$, thus,  the fit functions to the integral energy spectrum provides us the spectral index $ \alpha $.   We plot $\sum_k^{\infty} E_u(k')$ and $\Pi_u(k)$ versus $k$ in Fig.~\ref{fig:KESpectra_Flux}(a,b). The figure shows that For $\mathrm{Pr} \leq 1$,  $\sum_k^{\infty} E_u(k')$ scales  as $ k^{-2/3}$ for intermediate wavenumbers, which translates to Kolmogorov's energy spectrum  ($E_u(k) \sim k^{-5/3}$). The errors in the exponents obtained from the best fits range from $\pm 0.01$ to $\pm 0.07$. Further, consistent with the observed Kolmogorov's energy spectrum, $\Pi_u(k)$ is approximately constant over these wavenumbers. Our results, which are based on convection with no-slip walls, are consistent with earlier works~\cite{Mishra:PRE2010,Kumar:PRE2014,Kumar:PRE2015,Verma:NJP2017} on small and moderate Pr convection but with free-slip walls.  These observations rule out Bolgiano-Obukhov scaling ($ E_u(k) \sim k^{-11/5} $) for thermal convection.  Earlier, based on positive kinetic-energy injection rate by buoyancy,  \citet{Kumar:PRE2014} and \citet{Verma:NJP2017} had argued in favour of Kolmogorov's spectrum for turbulent convection.

For $\mathrm{Pr} = 6.8$ and $100$, the kinetic energy flux $\Pi_u(k)$ decreases sharply with $ k $ in the inertial range. Thus, instead of Kolmogorov's spectrum, we obtain a much steeper energy spectrum:  $\sum_k^{\infty} E_u(k') \sim k^{-2.49}$ for $\mathrm{Pr}=6.8$ and $\sim k^{-2.62}$ for $\mathrm{Pr}=100$, with an error of $\pm 0.01$ in the exponents. These relations translate to  $E_u(k) \sim k^{-3.49}$ for $\mathrm{Pr}=6.8$ and $\sim k^{-3.62}$ for $\mathrm{Pr}=100$.  For $\mathrm{Pr} \rightarrow \infty$, \citet{Pandey:PRE2014} derived that $ E_u(k) \sim k^{-13/3} $.  Note that the energy spectra for $\mathrm{Pr}=6.8$ and $100$ are quite close to the energy spectrum for $\mathrm{Pr} \rightarrow \infty$, consistent with the earlier results on energy spectra and fluxes~\cite{Pandey:PRE2014,Verma:book:BDF,Pandey:Pramana2016}.

Now, we explore the Pr dependence of the amplitudes of the kinetic energy spectra and fluxes. Figure~\ref{fig:KESpectra_Flux}(a,b) shows that for the same Ra,  convection with small Pr has more kinetic energy than that with large Pr. This is because  the nonlinear interactions among the velocity modes for small-Pr convection are stronger than those for large-Pr convection. {\color{black} Our results are in agreement with the earlier studies that report the Reynolds number (which indicates the strength of nonlinearities) to be a decreasing function of Pr~\cite{Ahlers:RMP2009,Verzicco:JFM1999,Lam:PRE2002,Silano:JFM2010}.} Further, for $\mathrm{Pr}\leq 1$, the width of the wavenumbers' range over which Kolmogorov's scaling is observed decreases with the increase of Pr:   $8 \leq k \leq 60$ for $\mathrm{Pr}=0.02$ and $6 \leq k \leq 17$ for $\mathrm{Pr}=1$. Note, however, that for large Prandtl numbers, power law regimes are observed at much larger wavenumbers.


Having analyzed the energy spectra and fluxes, we now examine the variations of the kinetic energy injection rates $\mathcal{F}_B(k)$ with Pr. We plot $\mathcal{F}_B(k)$ versus $k$ for different Prandtl numbers  in Fig.~\ref{fig:KESpectra_Flux}(c).
These plots reveal that $\mathcal{F}_B(k)$ is positive for all Pr, implying that buoyancy feeds kinetic energy to the system. This observation is in agreement with the findings of \citet{Kumar:PRE2014} and \citet{Verma:NJP2017} for $\mathrm{Pr}=1$ and contradicts the earlier arguments favoring BO scaling in RBC~\cite{Procaccia:PRL1989,Lvov:PRL1991,Lvov:PD1992,Rubinstein:NASA1994,Benzi:EPL1994,Benzi:EPL1994b,Ching:PRE2000,Ashkenazi:PRL1999spectrum,Shang:PRE2001,Calzavarini:PRE2002,Kunnen:PRE2008}. 

Figure~\ref{fig:KESpectra_Flux}(c) also shows that the kinetic energy injection is the strongest for $\mathrm{Pr}=0.02$ and becomes weaker as Pr increases, similar to the energy spectrum and flux. Further, for small Prandtl numbers, $\mathcal{F}_B(k)$ drops sharply with $k$ compared to larger Prandtl numbers. This is because, in the limit of $\mathrm{Pr} \rightarrow 0$, $\mathcal{F}_B(k)$ scales as $\mathrm{Ra} \langle |u_z (k)|^2 \rangle /k^2$~\cite{Mishra:PRE2010,Verma:book:BDF}, which shows that $\mathcal{F}_B(k)$' decreases steeply with $k$. Thus, for small and moderate Prandtl numbers, $\mathcal{F}_B(k)$ is small in the inertial range compared to the energy flux and cannot 
bring significant variations in $\Pi_u(k)$ in that regime. This results in Kolmogorov-like scaling of the kinetic energy spectrum for small and moderate-Pr convection, consistent with the arguments of earlier studies~\cite{Mishra:PRE2010,Kumar:PRE2014,Verma:NJP2017,Verma:PS2019}.



In Fig.~\ref{fig:KESpectra_Flux}(d), we plot the normalized derivative of the kinetic energy flux, $d_k \Pi_u/\Pi_u(k)$, versus $k$ for different Pr ($d_k$ denotes the derivative with respect to $k$). Recall from Eq.~(\ref{eq:VEF_RBC}) that $d_k \Pi_u(k) = \mathcal{F}_B(k)-D(k)$. Since energy is dissipated at small scales, $D(k)$ becomes stronger than $\mathcal{F}_B(k)$ at large wavenumbers, causing the kinetic energy flux to be a decreasing function of $k$. As evident in Fig.~\ref{fig:KESpectra_Flux}(d), the crossover wavenumber at which the derivative of the flux changes sign decreases with increasing Pr: $k=33$ for $\mathrm{Pr}=0.02$ and $k=14$ for $\mathrm{Pr}=100$. This is expected; since Ra is constant in all the runs, the flow is more viscous and less thermally diffusive by the same factor for increasing Pr. Hence, $D(k)$ is strong even at intermediate scales~\cite{Pandey:PRE2014}.  For $\mathrm{Pr}=6.8$ and $100$, $D(k)$ exceeds $\mathcal{F}_B(k)$ by a significant amount at intermediate scales, resulting in $d_k \Pi_u(k)/\Pi_u(k)  \lesssim -0.1$. Thus, $\Pi_u(k)$ decreases sharply with $k$ for these Prandtl numbers in the intermediate scales, leading to a steeper energy spectrum compared to $k^{-5/3}$, consistent with the findings of \citet{Pandey:PRE2014,Pandey:Pramana2016}.

These results  provide a comprehensive picture for the variations of   kinetic energy spectra and fluxes of thermal convection with Pr. In the next subsection, we discuss  how the strength of the nonlinear interactions in RBC vary with Pr.

\subsection{Energy flux and viscous dissipation in thermal convection}
\label{subsec:Injection_dissipation}

In 3D hydrodynamic turbulence, the kinetic energy flux in the inertial range matches with the total dissipation rate.  This is not the case in turbulent thermal convection because buoyancy feeds energy at all scales, including the dissipation range.  Consequently, $ \Pi_u < \epsilon_u $. \citet{Bhattacharya:PF2019b} showed that for Pr = 1, the inertial-range kinetic energy flux is approximately one-third of the total dissipation rate. In this subsection, we will describe these quantities for various Prandtl numbers.

\begin{figure}[t]
	\includegraphics[scale=0.25]{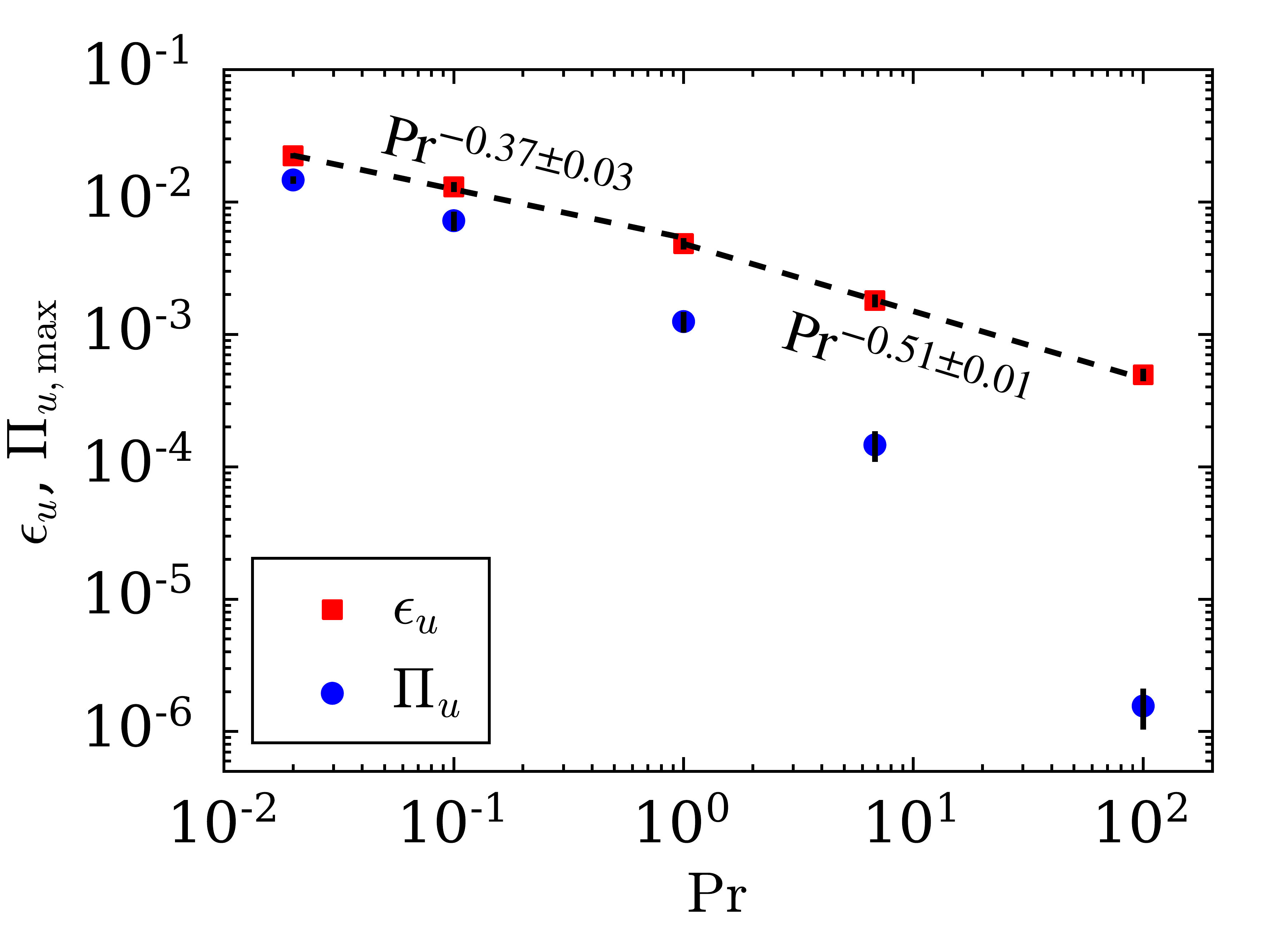}
	\caption{(color online) For $\mathrm{Ra}=10^7$: Plots of $\epsilon_u$ and maximum kinetic energy flux $\Pi_{u,\mathrm{max}}$ vs. $k$. The dissipation rates decrease with the increase of Pr, similar to the energy spectrum and flux. The difference between the kinetic energy flux and the dissipation rate increases as Pr is increased.} 
	\label{fig:Flux_Dissipation}
\end{figure}

In Fig.~\ref{fig:Flux_Dissipation}, we plot the total viscous dissipation rate ($\epsilon_u$) along with the maximum inertial range kinetic energy flux ($\Pi_{u,\mathrm{max}}$) versus Pr. {\color{black}The figure shows that $\epsilon_u$ decreases with the increase of Pr as $\mathrm{Pr}^{-0.37}$ for $\mathrm{Pr} < 1$, and as $\mathrm{Pr}^{-0.51}$  for $\mathrm{Pr} \geq 1$. Our observations are consistent with the fact that the nonlinear interactions among the velocity modes decrease with increasing Pr. It is also clear from Fig.~\ref{fig:Flux_Dissipation} that $ \Pi_{u,\mathrm{max}} < \epsilon_u $ as discussed earlier. Further, the difference between $ \Pi_{u,\mathrm{max}}$ and $\epsilon_u$ increases as Pr is increased. Recall from Sec.~\ref{subsec:KESpectrum} that the kinetic energy injection rate $\mathcal{F}_B(k)$ decreases steeply with $k$ for small Pr and becomes progressively less steep as Pr is increased [as exhibitted in Fig.~\ref{fig:KESpectra_Flux}(c)]. This indicates that for small Pr, most of the energy is injected at large scales, as a result of which the inertial range kinetic energy flux is only marginally less than the total dissipation rate. Thus, small Pr convection is close to 3D hydrodynamic turbulence where $\Pi_u \approx \epsilon_u$. On the other hand, for large Pr, }only a small fraction of the total energy is injected at large scales and a significant amount of kinetic energy is injected in the inertial and dissipation ranges. Therefore, the inertial-range flux is much less than $\epsilon_u$.   For $\mathrm{Pr}=100$, the inertial-range kinetic energy flux is about three orders of magnitude smaller than the dissipation rate. 

\begin{figure}[t]
	\includegraphics[scale=0.25]{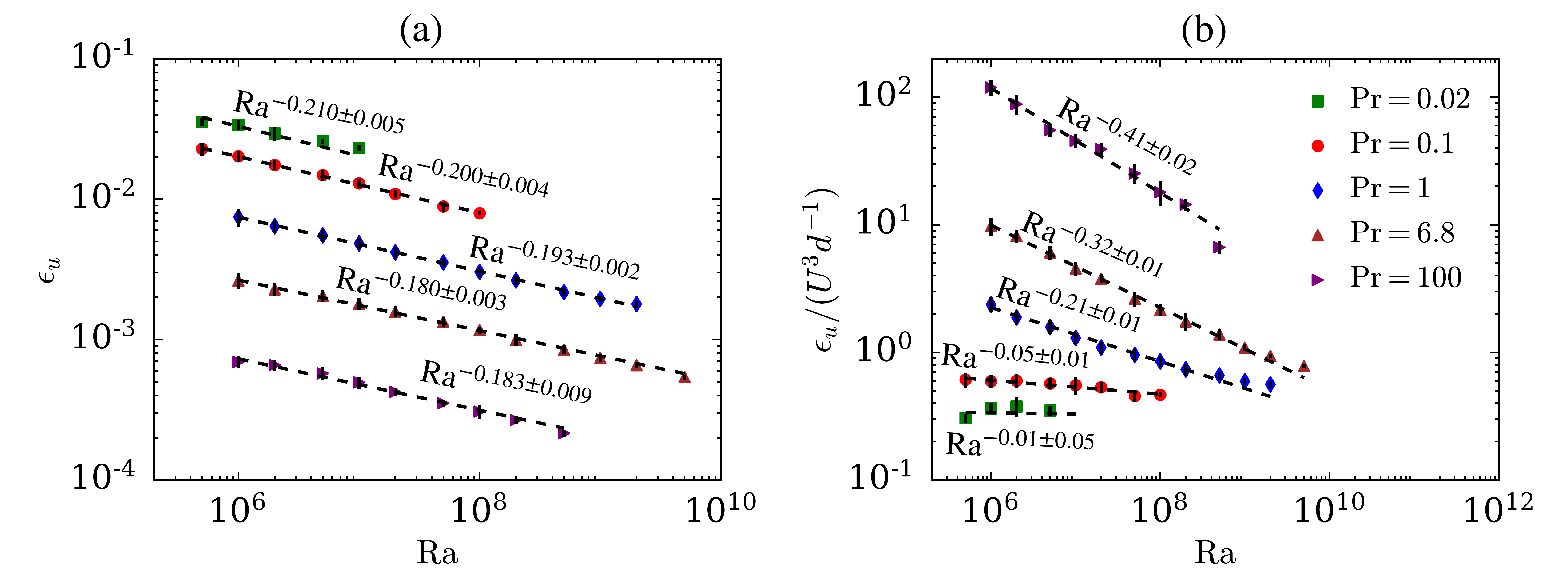}
	\caption{(color online) For $\mathrm{Pr}=0.02$, $0.1$, $1$, $6.8$ , and $100$, {\color{black}(a) plots of $\epsilon_u$ vs. Ra}, and (b) plots of $\epsilon_u/(U^3/d)$ vs. Ra (data taken from \citet{Bhattacharya:PF2020}). For small Pr, $\epsilon_u \sim U^3/d$ as in hydrodynamic turbulence. However, $\epsilon_u$ has an additional Ra dependence for larger Prandtl numbers.} 
	\label{fig:Eu_Ra}
\end{figure}

Now, we compare the scaling of $\epsilon_u$ in RBC and hydrodynamic turbulence. For the latter,    
\begin{equation}
\epsilon_u \approx U^3/d, \nonumber
\end{equation}
where $U$ is the large scale velocity (for example, the root mean square velocity), and $d$ is the size of the domain. 
However, in thermal convection, \citet{Pandey:PF2016} and \citet{Pandey:PRE2016} showed that for $\mathrm{Pr}=1$,
\begin{equation}
\epsilon_u \sim (U^3/d) \mathrm{Ra}^{-0.18}, \nonumber
\end{equation} 
instead of $U^3/d$; here $U$ is the root mean square velocity. The additional Ra dependence was attributed to multiscale forcing by buoyancy and to the suppression of nonlinear interactions due to the presence of walls. Motivated by these observations, we investigate the scaling of viscous dissipation rate for various  Prandtl numbers. Towards this objective, we use additional datasets of \citet{Bhattacharya:PF2020}  that include simulations for Ra ranging from $5 \times 10^5$ to $2 \times 10^9$  and Pr ranging from $0.02$ to $100$.  {\color{black}In Fig.~\ref{fig:Eu_Ra}(a), we plot $\epsilon_u$ versus Ra.} We compute $\epsilon_u/(U^3/d)$ for all the  data points and plot them versus Ra in Fig.~\ref{fig:Eu_Ra}(b). We also plot the best-fit curves for our data on {\color{black}the above figures.} 

{\color{black} Figure~\ref{fig:Eu_Ra}(a) shows that the viscous dissipation rate $\epsilon_u$ decreases with Ra as $\mathrm{Ra}^{-\alpha}$, where $\alpha$ ranges from $0.21$ for $\mathrm{Pr}=0.02$ to $0.18$ for $\mathrm{Pr}=6.8$ and $100$. Our results are consistent with those of \citet{Scheel:PRF2017}, who also observed a similar value for the exponent $\alpha$ and its variations with Pr. Figure~\ref{fig:Eu_Ra}(b)} shows that for $\mathrm{Pr}=0.02$, {\color{black}$\epsilon_u \sim U^3/d$, similar to} hydrodynamic turbulence.   However, for larger Pr, $\epsilon_u/(U^3/d)$ decreases with Ra with  slopes  getting steeper with the increase of Pr.  For $\mathrm{Pr} = 100$, {\color{black}$\epsilon_u$ has a strong Ra correction with $\epsilon_u \sim (U^3/d)\mathrm{Ra}^{-0.41}$.}  The strong Ra dependence for large Pr is due to strong viscous dissipation in such flows.


In the next section, we discuss the Pr dependence on the velocity structure functions of turbulent convection.

\section{Structure functions}
\label{sec:SFs}
 In this section, we examine the velocity structure functions of turbulent convection using our numerical data.
  
 For   homogeneous and isotropic 3D hydrodynamic turbulence, \citet{Kolmogorov:DANS1941Dissipation,Kolmogorov:DANS1941Structure} proved the following exact relation for the third-order structure function:
\begin{equation}
S_3^u(l) = - \frac{4}{5} \epsilon_u l \sim -l.
\label{eq:Kolmogorov_SF}
\end{equation}
However, for $q \ne 3$, the velocity structure functions scale as $S_q^u(l) \sim \pm l^{\zeta_q}$, where $\zeta_q$ is  the exponents for $ q$-th order structure function.  Note that a simple extrapolation of Kolmogorov's theory to higher-order structure functions yields $ \zeta_q = q/3 $ (labelled as KO41 model).  Several experimental and numerical results however do not agree with this prediction.  Among many models, such as $ \beta $ model, multifractal model, log-normal model~\cite{Frisch:book,Sreenivasan:ARFM1997,Verma:arxiv2005}, the model by \citet{She:PRL1994} provides the best fit function to $\zeta_q$, which is
\begin{equation}
\zeta_q = \frac{q}{9} + 2\left(1 - \left( \frac{2}{3} \right)^{q/3} \right).
\label{eq:She1994}
\end{equation}
This model is labelled as SL94.   The differences between the KO41 and SL94 are attributed to the {\em intermittency effects}~\cite{Frisch:book,Sreenivasan:ARFM1997}.

For turbulent thermal convection, some researchers have argued in favour of K041 model, while some others have argued that $\zeta_q = 3q/5 $, which is derived from Bolgiano-Obukhov scaling (labelled as BO59). For example, \citet{Benzi:EPL1994,Benzi:EPL1994b} computed the structure functions upto sixth order using  numerical data and reported BO59 scaling of $S_q^u(l) \sim l^{3q/5}$. \citet{Ching:PRE2000}, \citet{Calzavarini:PRE2002}, and \citet{Kunnen:PRE2008} also observed BO59 scaling. On the other hand, \citet{Sun:PRL2006}, and \citet{Kaczorowski:JFM2013} reported KO41 scaling for the lower order structure functions of RBC. 
Recently,  \citet{Bhattacharya:PF2019b} showed that the third-order velocity structure function of RBC for $\mathrm{Pr} = 1$  obeys 
Eq.~(\ref{eq:Kolmogorov_SF}), similar to hydrodynamic turbulence, except that $\epsilon_u$ is replaced with $\Pi_u$.  \citet{Sun:PRL2006} and \citet{Bhattacharya:PF2019b} showed that the exponents for the higher order structure functions of convection follow She-Leveque's model given by Eq.~(\ref{eq:She1994}).

\begin{figure}[b]
 	\includegraphics[scale=0.26]{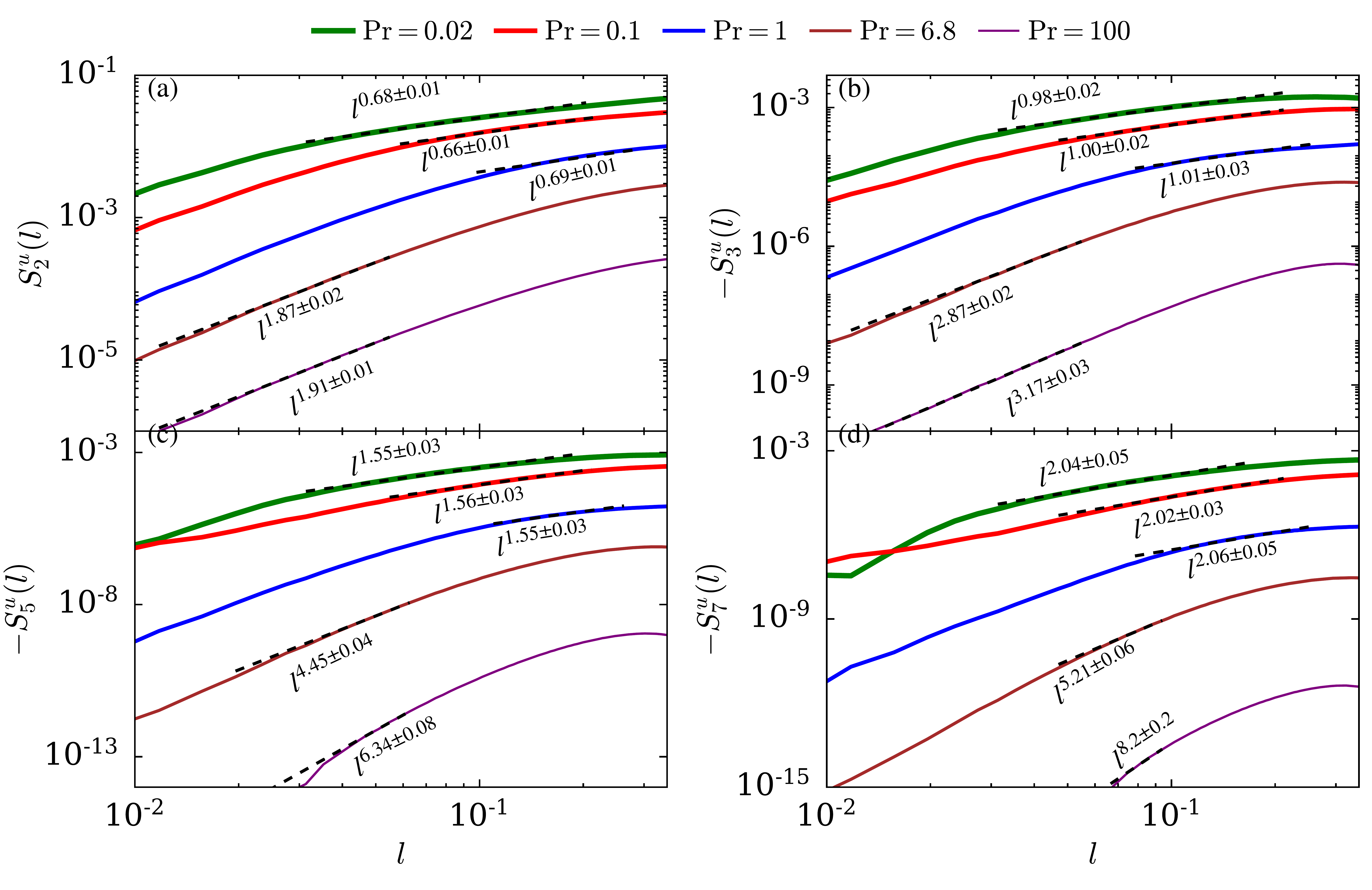}
 	\caption{(color online) For $\mathrm{Pr}=0.02$, $0.1$, $1$, $6.8$, and $100$: longitudinal velocity structure functions of orders (a) 2, (b) 3, (c) 5, and (d) 7 vs. $l$. The amplitudes of the structure functions decrease with the increase of Pr.} 
 	\label{fig:SFs}
 \end{figure}

\begin{figure}[t]
 	\includegraphics[scale=0.225]{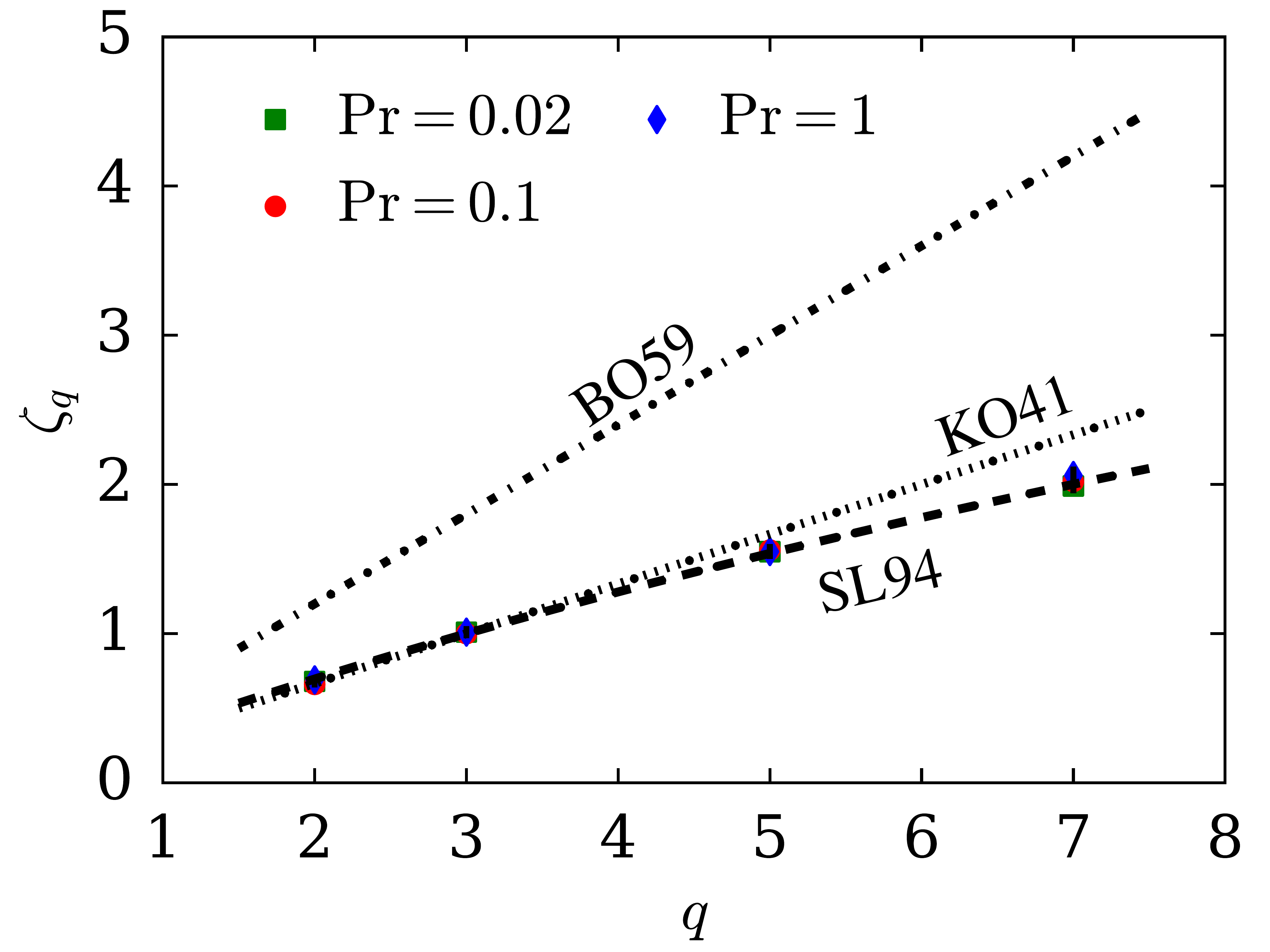}
 	\caption{(color online) For $\mathrm{Pr}=0.02$, $0.1$, and $1$: The scaling exponents $\zeta_q$ for the velocity structure functions vs. order $q$. The exponents match closely with the predictions of \cite{She:PRL1994} (SL94). The figure also exhibits the extrapolated $\zeta_q$'s for  K41 ($ q/3 $) and Bolgiano-Obukhov (BO59)  ($ 3q/5 $) .} 
 	\label{fig:She-Leveque}
 \end{figure}

Note that the above works are for a specific set of parameters, and they do not provide us with a comprehensive picture under the variations of Pr.  In the following discussion we  examine the  scaling as well as the relative strengths of velocity structure functions for Pr ranging from 0.02 to 100.  We compute the second, third, fifth, and seventh-order velocity structure functions using our numerical data (see Eq.~(\ref{eq:SF_q})).
We plot the second-order velocity structure function $S_2^u(l)$ versus $l$ in Fig.~\ref{fig:SFs}(a), and the negative of third, fifth, and seventh-order velocity structure functions [$-S_3^u(l)$, $-S_5^u(l)$, $-S_7^u(l)$] versus $l$ in Fig.~\ref{fig:SFs}(b,c,d). We also plot the respective best-fit curves in the same figures. We observe that for $\mathrm{Pr} \lessapprox 1$, the third-order structure function exhibits Kolmogorov's scaling of $S_3^u(l) \sim -l$ over a range of intermediate scales that  corresponds to the inertial range over which $E_u(k) \sim k^{-5/3}$ (as reported in Sec.~\ref{subsec:KESpectrum}).  In addition, for the above Prandtl numbers, the structure functions of orders $q=2$, $5$, and $7$ follow the predictions of \citet{She:PRL1994} [see Figs.~\ref{fig:SFs}(a,c,d) and \ref{fig:She-Leveque}], as in hydrodynamic turbulence. The errors in the exponents range from $\pm 0.01$ for the third-order structure functions to $\pm 0.05$ for the seventh-order structure functions. Our results are thus consistent with those of \citet{Sun:PRL2006} and \citet{Bhattacharya:PF2019b}, but contrary to the studies that report Bolgiano-Obukhov scaling of the structure functions~\cite{Benzi:EPL1994,Benzi:EPL1994b,Ching:PRE2000,Calzavarini:PRE2002,Kunnen:PRE2008}.

Figure~\ref{fig:SFs}(a,b,c,d) also shows that the amplitudes of the velocity structure functions for all orders increase with decreasing Pr, similar to the amplitudes of the kinetic energy spectrum and flux. This is expected because the structure functions are directly related to the kinetic energy spectrum and flux~\cite{Ching:book,Lesieur:book:Turbulence,Frisch:book}, which show similar scaling (see Sec.~\ref{sec:Results}).




The structure functions for $\mathrm{Pr}=6.8$ and $100$ neither follow Kolmogorov's scaling  nor She-Leveque's scaling;  instead, they vary steeply at intermediate scales compared to those for $\mathrm{Pr} \leq 1$ (see Fig.~\ref{fig:SFs}). 
For $\mathrm{Pr}=6.8$, the structure functions scale as {\color{black}$S_2^u(l) \sim l^{1.87 \pm 0.02}$, $S_3^u(l) \sim -l^{2.87 \pm 0.02}$, $S_5^u(l) \sim -l^{4.45 \pm 0.04}$, and $S_7^u(l) \sim -l^{5.21 \pm 0.06}$.}
For $\mathrm{Pr}=100$, the curves are even steeper, with the structure functions scaling as {\color{black}$S_2^u(l) \sim l^{1.91 \pm 0.01}$, $S_3^u(l) \sim -l^{3.17 \pm 0.03}$, $S_5^u(l) \sim -l^{6.34 \pm 0.08}$, and $S_7^u(l) \sim -l^{8.2 \pm 0.2}$.} 
Recall that in the limit of infinite Pr, the energy spectrum scales as $E_u(k) \sim k^{-13/3}$ due to strong viscous dissipation in the intermediate scales.  A simple extrapolation of the above to the second, third, fifth, and seventh-order structure functions lead to $S_2^u(l) \sim l^{10/3}$,  $S_3^u(l) \sim -l^{5}$,  $S_5^u(l) \sim -l^{25/3}$, and  $S_2^u(l) \sim -l^{35/3}$ respectively (without intermittency effects~\cite{Frisch:book}). The slopes of the structure functions computed using our data for $\mathrm{Pr}=6.8$ and $100$ are not as steep as above predictions; this is possibly because the Prandtl numbers for our runs are finite and there are possible intermittency effects. Nevertheless, it is evident that the slopes of the structure functions for larger Prandtl numbers are significantly steeper than those for smaller Prandtl numbers. 

In the next section, we discuss the Prandtl number effects on the probability distribution functions of convective heat flux.

\section{Prandtl number dependence of local heat flux}
\label{sec:PDFs}

There is a net heat transport in thermal convection, which is quantified using a nondimensional number called Nusselt number (Nu):
\begin{equation}
\mathrm{Nu} = 1 + \frac{\langle u_zT \rangle}{\kappa \Delta /d}, \label{eq:Nu}
\end{equation}
where $T$ is the temperature field, and $ u_z $ is the vertical velocity.   The Nusselt number is always positive, but the local vertical heat flux, given by $u_zT$, exhibits strong fluctuations~\cite{Shang:PRL2003,Kaczorowski:JFM2013,Shishkina:PF2007,Pharasi:PF2016}.  It has been observed that $u_zT$ take both positive and negative values, but  the positive $u_zT$ dominates the negative ones leading to a net vertical heat flux. {\color{black}The strength of the fluctuations further vary in different regions inside the RBC domain; the fluctuations are strongest near the sidewalls~\cite{Shang:PRL2003} and weakest near the horizontal walls~\cite{Shishkina:PF2007}}. In this section, we present the variations of the probability distribution function (PDF) of $u_zT$  with the Prandtl number.  In addition, we also study the horizontal heat fluxes, $u_xT$ and $u_yT$, which are expected to be symmetric so as to yield a zero net flux along the horizontal directions. 
 \begin{figure}[b]
	\includegraphics[scale=0.6]{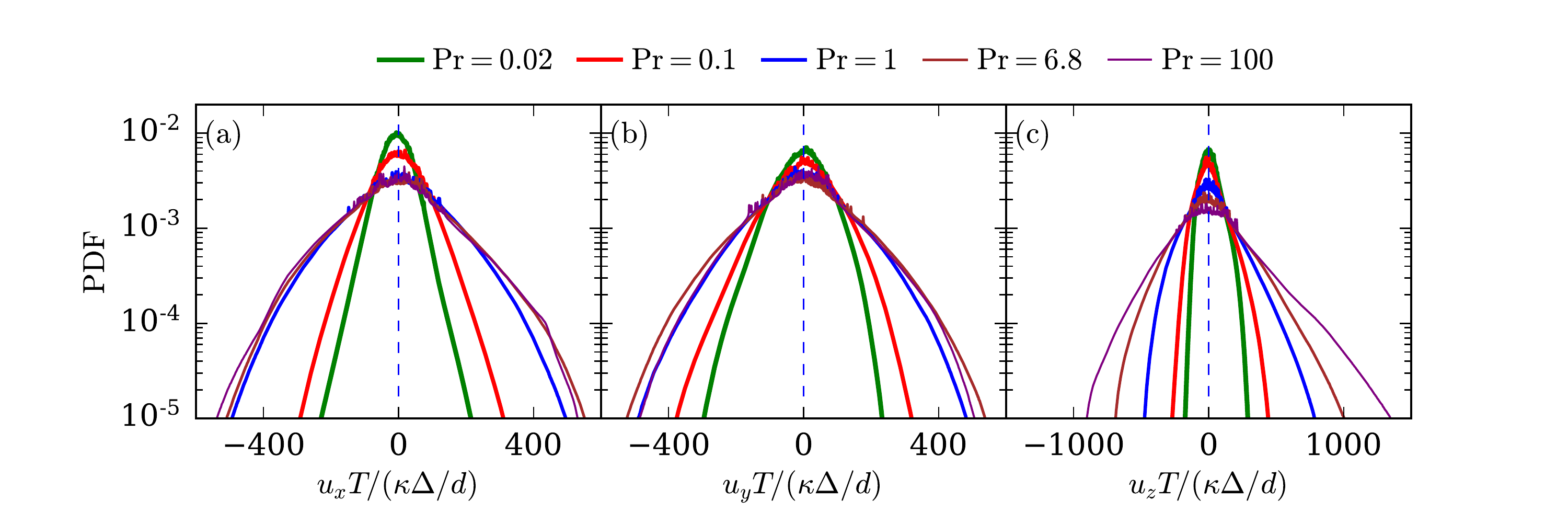}
	\caption{(color online) For $\mathrm{Ra}=10^7$: The probability distribution functions (PDFs) of normalized local convective heat flux in the (a) $x$ direction, (b) $y$ direction, and (c) $z$ direction for different Pr. The fluctuations of the local heat flux increase with Pr.}
	\label{fig:PDF_1}
\end{figure}
\begin{figure}[t]
	\includegraphics[scale=0.6]{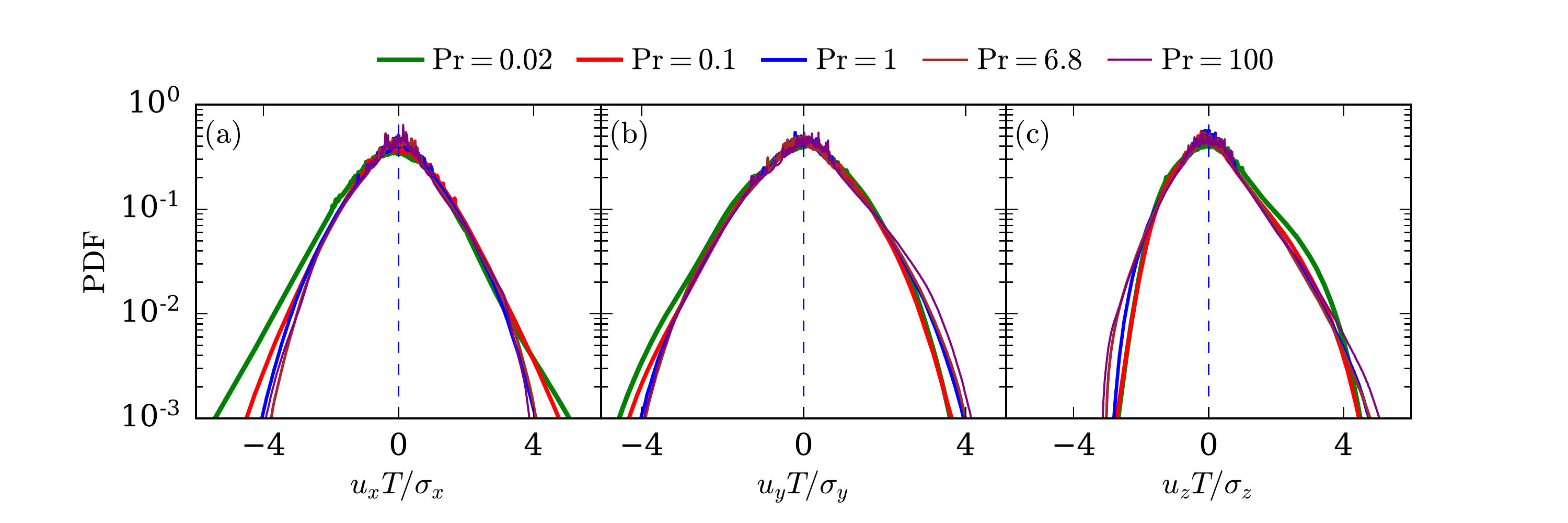}
	\caption{(color online) For $\mathrm{Ra}=10^7$: The probability distribution functions (PDFs) of the local convective heat flux normalized with their respective standard deviations ($\sigma_x$, $\sigma_y$, $\sigma_z$) in the (a) $x$ direction, (b) $y$ direction, and (c) $z$ direction. The normalized PDFs for different Prandtl numbers collapse  into one curve.}
	\label{fig:PDF_2}
\end{figure}	


We compute the PDFs of the heat fluxes {\color{black}over the entire domain} using our simulation data. The PDFs are averaged over 31 to 101 timeframes depending on the Prandtl numbers (see Table~\ref{table:PDFs}). We plot the PDFs of the horizontal heat fluxes $u_xT$ and $u_yT$, normalized by $\kappa \Delta/d$, in Fig.~\ref{fig:PDF_1}(a,b) respectively, and the vertical heat flux $u_zT$,  normalized by $\kappa \Delta/d$, in Fig.~\ref{fig:PDF_1}(c). Note that $\kappa=(\mathrm{RaPr})^{-1/2}$ from Eq.~(\ref{eq:NDTheta}). For all Prandtl numbers, the horizontal and vertical heat fluxes peak at zero. However,  the horizontal heat fluxes are symmetric about their peaks, but the vertical heat fluxes show clear asymmetry with long tails in the positive direction.  The asymmetry in the vertical flux yields a net vertical heat transport, but the symmetric horizontal fluxes sum to zero, as expected.   These results are consistent with earlier studies~\cite{Pharasi:PF2016,Shang:PRL2003,Kaczorowski:JFM2013,Shishkina:PF2007}.  Both horizontal and vertical heat fluxes exhibit strong fluctuations near their most probable value of zero, causing noise-like ripples near the peaks of their PDFs.


\begin{table}
	\caption{For $\mathrm{Pr}=0.02$ to $100$ and $\mathrm{Ra}=10^7$: Standard deviations ($\sigma_x$, $\sigma_y$, and $\sigma_z$) of the local heat fluxes $u_xT$, $u_y T$, and $u_z T$ respectively, and the number of snapshots over which the PDFs of the above quantites are averaged. The standard deviations increase with Pr.}
	\begin{ruledtabular}
		\begin{tabular}{c c c c c}
			$\mathrm{Pr}$ &  $\sigma_x$ & $\sigma_y$ & $\sigma_z$ & Snapshots\\
			\hline 
			$0.02$ & {\color{black} $36 \pm 1$} & $62 \pm 1$ & {\color{black}$64 \pm 1$} & {\color{black}81}\\ 
			$0.1$ & $62 \pm 5$ & $87 \pm 8$ & $99 \pm 7$ & {\color{black}66}\\
			$1$ &  $125 \pm 11$ & $125 \pm 11$ & $173 \pm 7$ & 101\\
			$6.8$ &  $139 \pm 16$ & $138 \pm 14$ & $231 \pm 13$ & 101\\
			$100$ &  $140 \pm 18$ & $124 \pm 16$ & $296 \pm 21$ & 101
		\end{tabular}
		\label{table:PDFs}
	\end{ruledtabular}
\end{table}

A careful observation of the PDFs of Fig.~\ref{fig:PDF_1} show an interesting feature:  the fluctuations in the local heat fluxes increase with the Prandtl numbers, which  is evident from the long tails for $\mathrm{Pr} \geq 1$.   {\color{black}It is to be noted that for the same Rayleigh number, the thermal diffusivity, which is given by $\kappa = 1/\sqrt{\mathrm{RaPr}}$, is strong for small Prandtl numbers. Hence, due to high thermal diffusivity, the heat transport is primarily via diffusion for small Pr, resulting in thick thermal plumes. The high value of thermal diffusivity and the thick plumes result in weak fluctuations of the local heat flux for small Pr.}  However, large-Pr convection takes place via thin thermal plumes {\color{black}due to weak thermal diffusivity, thereby} inducing strong thermal fluctuations and inhomgeneity in the heat flux~\cite{Silano:JFM2010,Pandey:PRE2014,Pandey:Pramana2016}.

Interestingly, the PDFs of Fig.~\ref{fig:PDF_1}(a,b,c) can be collapsed into one curve each by normalizing the curves using the corresponding standard deviations. We present the collapsed curves in Fig.~\ref{fig:PDF_2}(a,b,c).  The standard deviations are computed for every timeframe and then averaged. The computed standard deviations for different Prandtl numbers are tabulated in Table~\ref{table:PDFs}.  As expected, the standard deviations along with their respective errors increase with Prandtl number. There is, however, an anomaly in $\sigma_y$ for $\mathrm{Pr}=100$ in that it is less than that for $\mathrm{Pr}=6.8$. However, we believe that this is a minor aberration that can be resolved by averaging over more timeframes.


We conclude in the next section.

\section{Summary and conclusions}
\label{sec:Conclusions}



In this paper, using detailed numerical simulations of turbulent convection, we analyzed the Prandtl number dependence of the kinetic energy spectrum, flux, and the spectra of buoyant energy injection and viscous dissipation rates. Additionally, we examined the variations of velocity structure functions and the local heat flux with Pr. For our analysis, we varied Pr from 0.02 to 100, keeping the Rayleigh number fixed at $\mathrm{Ra}=10^7$. 

Consistent with earlier works, the kinetic energy spectrum  exhibits Kolmogorov scaling of $\sim k^{-5/3}$ for $\mathrm{Pr}\leq 1$ and a steeper scaling of {\color{black}$\sim k^{-2.5}$} for $\mathrm{Pr}\gg 1$~\cite{Mishra:PRE2010,Kumar:PRE2014,Kumar:PRE2015,Verma:NJP2017,Kumar:RSOS2018,Pandey:PRE2014,Pandey:Pramana2016}.  The inertial range is widest for $\mathrm{Pr}=0.02$, and it gets narrower as Pr is increased.
The magnitudes of the kinetic energy spectrum and flux decrease with Pr, implying that flows with small Prandtl number have stronger nonlinear interactions among the velocity modes. 
The amplitudes of kinetic energy injection and dissipation rates follow a similar pattern as  energy flux and spectrum. {\color{black}Our results are in agreement with earlier studies that report the Reynolds number to be a decreasing function of Pr~\cite{Verzicco:JFM1999,Lam:PRE2002,Silano:JFM2010,Ahlers:RMP2009}.} For $\mathrm{Pr} \ll 1$, kinetic energy injection by buoyancy occurs mostly at large scales, causing the kinetic energy flux in the inertial range to be approximately equal to the viscous dissipation rate, similar to hydrodynamic turbulence. On the other hand, for $\mathrm{Pr} \gg 1$, significant kinetic energy is injected at small scales as well, causing the energy flux to be a small fraction of the viscous dissipation rate.

The amplitudes of the velocity structure functions increase with the decrease of Pr, consistent with the results on energy spectrum. The velocity structure functions for $\mathrm{Pr} \leq 1$ were shown to be in agreement with She-Leveque's model, similar to hydrodynamic turbulence and consistent with earlier results~\cite{Bhattacharya:PF2019b,Sun:PRL2006}. The structure functions exhibit steeper curves for $\mathrm{Pr}=6.8$ and $100$ and are in agreement with the scaling of the energy spectrum  for  large Prandtl numbers. 

The  strength of fluctuations of the local convective heat flux increases with  Pr.  This is because the thick thermal plumes for small-Pr flows transfer heat efficiently throughout the flow, but thin thermal plumes for large-Pr flows create strong inhomogeneity in the heat flux.

 Thus, our present study provides valuable insights into the  variations of turbulent velocity and thermal fluctuations  with Pr. Although we worked on a small set of parameters, we expect these patterns to be valid over a wide range or Ra and Pr, with the possible exception of the ultimate regime~\cite{Kraichnan:PF1962Convection}. We expect these results to be important for modeling flows  in stars, bubbly turbulence, and liquid metal batteries. Further, our analysis should also help in developing accurate subgrid models for convection.  

\section*{Acknowledgements}
The authors are grateful to Roshan Samuel and Syed Fahad Anwer for their important contributions in the development of the finite difference code SARAS. Further, the authors thank Shadab Alam and Soumyadeep Chatterjee for useful discussions. Our numerical simulations were performed on Shaheen II of {\sc Kaust} supercomputing laboratory, Saudi Arabia (under the project k1416) and on HPC2013 of IIT Kanpur, India. 
\appendix
\section{Entropy spectra and flux of RBC}
\label{sec:Entropy}
In this section, we compute the nondimensionalized  entropy spectra, $E_\theta(k)$, and entropy fluxes, $\Pi_\theta(k)$, of RBC using our data for different Pr, with $\mathrm{Ra}=10^7$. These quantities are computed as follows:
\bea
E_\theta(k) &=& \frac{1}{2} \sum_{k\leq |\mathbf{k}'| < k+1} |\theta(\mathbf{k}')|^2, 
 \label{eq:Entspectrum}\\
\Pi_\theta(k_0) &=& \sum_{k \geq k_0} \sum_{p<k_0} \delta_\mathbf{k,p+q} \Im(\mathbf{[k \cdot u(q)][\theta^*(k)  \theta(p)]}).
\label{eq:Eflux_MtoM} 
 \eea
 
\begin{figure}[b] 
 	\includegraphics[scale=0.25]{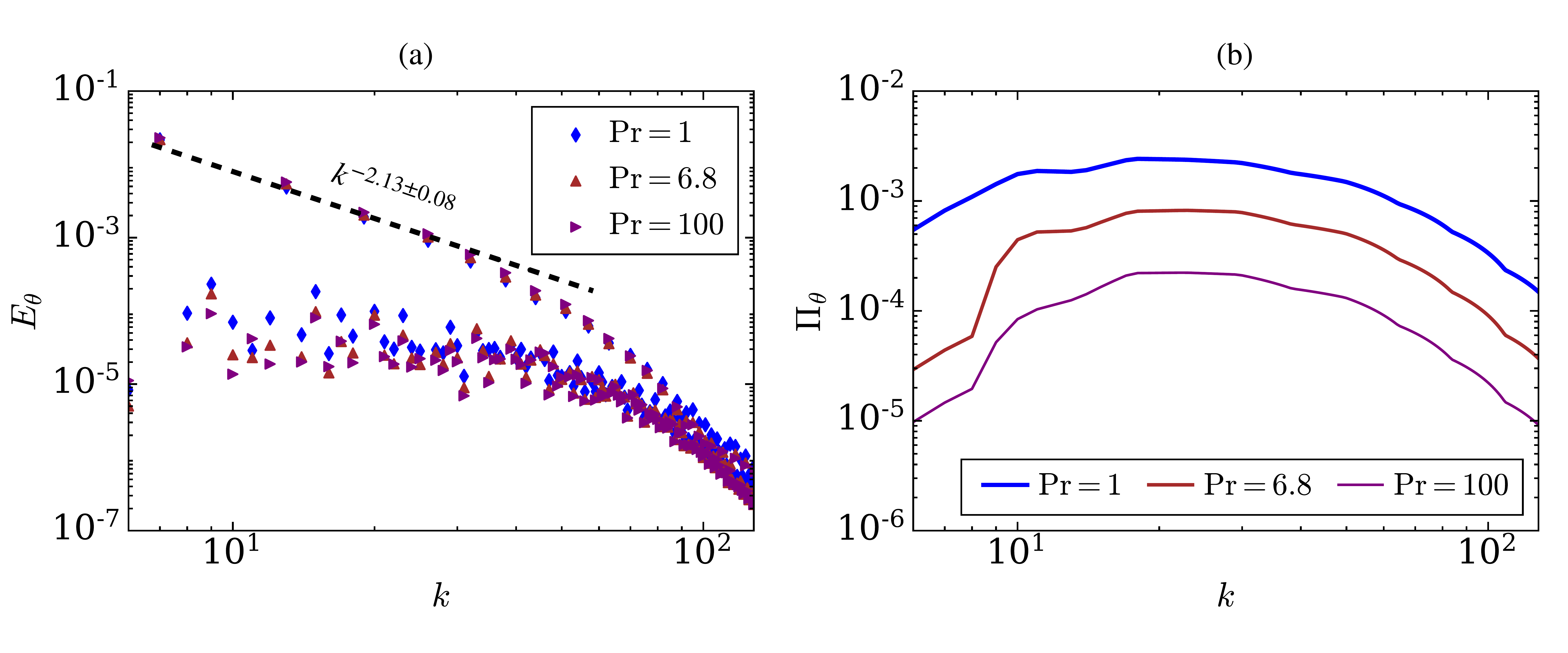}
 	\caption{(color online) For $\mathrm{Ra}=10^7$ and $\mathrm{Pr}=1$, $6.8$, and $100$: (a) Entropy spectrum $E_\theta$ (with dual branches) and (b) entropy flux $\Pi_\theta$ vs. $k$. The amplitudes of the entropy spectrum do not vary with Pr, but the amplitudes of the entropy flux decrease with increase of Pr.} 
 	\label{fig:Entropy_largePr}
 \end{figure}

\begin{figure}[htbp]
 	\includegraphics[scale=0.25]{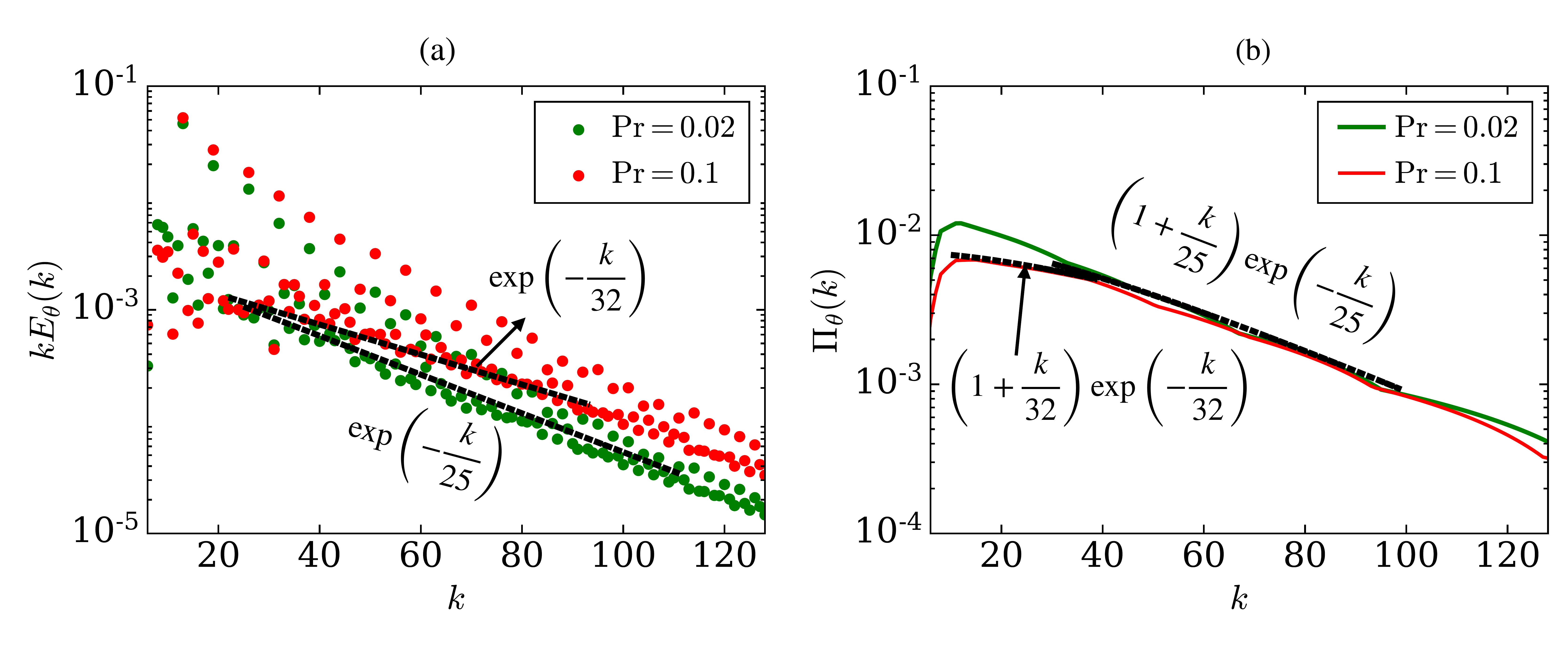}
 	\caption{(color online) For $\mathrm{Ra}=10^7$ and $\mathrm{Pr}=0.02$ and $0.1$: Semi-log plots of (a) normalized entropy spectrum $kE_\theta$ and (b) entropy flux $\Pi_\theta$ vs. $k$. The lower branch of the entropy spectrum and the entropy flux fit well with exponential function.} 
 	\label{fig:Entropy_smallPr}
 \end{figure}

We plot the entropy spectra and fluxes for $\mathrm{Pr} = 1$, $6.8$, and $100$ in Fig.~\ref{fig:Entropy_largePr}(a,b), and for $\mathrm{Pr} = 0.02$ and $0.1$ in Fig.~\ref{fig:Entropy_smallPr}(a,b).
The figures show that the nondimensional entropy are approximately the same for all Prandtl numbers, unlike the kinetic energy spectrum that decreases with the increase of Pr. The entropy flux, however, decreases with the increase of Pr because the entropy flux is proportional to the velocity  fluctuations  (see Eq.~(\ref{eq:Eflux_MtoM})), which are strong   for flows with small Pr. 


The entropy spectrum exhibits dual branch for $\mathrm{Pr} = 1$, $6.8$, and $100$, with the upper branch scaling as $\sim k^{-2.13 \pm 0.08}$.  
\citet{Mishra:PRE2010} and \citet{Pandey:PRE2014} explained this branch in terms of  the temperature modes $\theta(0,0,2n)$, which are approximately equal to $-1/(2n\pi)$ for thin thermal boundary layers ($n$ being an integer). The lower branch, which is constituted by the remaining modes, does not exhibit any clear scaling. The temperature modes of both the branches yield the constant entropy flux   (see Fig.~\ref{fig:Entropy_largePr}(b)). 

For $\mathrm{Pr}=0.02$ and $0.1$, the entropy spectrum again has two branches; however, the upper branch is not very prominent because of thick thermal boundary layers. For small-Pr convection, the nonlinear term of the $\theta$-equation [Eq.~(\ref{eq:theta})] is small compared to the diffusive term, similar to the momentum equation for laminar flows. Following the arguments of \citet{Martinez:JPP1997} and \citet{Verma:FD2018} for energy spectrum of laminar flows, we propose that the entropy spectrum for small-Pr convection is of the following exponential form:
\begin{equation}
E_\theta(k) \sim  k^{-1} (k/k_c) \exp(-k/k_c),
\label{eq:Entropy_smallPr_proposal}
\end{equation}
where $k_c$ is the wavenumber beyond which the thermal energy dissipation becomes dominant. 

Now, for a steady state, the entropy flux is related to entropy injection ($\mathcal{F}_\theta$) and dissipation spectra ($2 \kappa k^2 E_\theta$)  by the variable entropy flux equation:
\begin{equation}
\frac{d \Pi_\theta}{dk} = \mathcal{F}_\theta (k) - 2 \kappa k^2 E_\theta(k),
\label{eq:VEtF}
\end{equation}
In the intermediate wavenumbers for small-Pr convection, the spectrum of entropy dissipation dominates that of the entropy injection rate; hence $2\kappa k^2 E_\theta(k) \gg \mathcal{F}_\theta (k)$. Using this condition and substituting the expression of Eq.~(\ref{eq:Entropy_smallPr_proposal}) in Eq.~(\ref{eq:VEtF}), we obtain the following:
\begin{equation}
\frac{d \Pi_\theta}{dk}  \sim  k \exp(-k/k_c).
\label{eq:VEtF_Dissipation}
\end{equation}
Integration of the above expression yields the following expression for the entropy flux:
\begin{equation}
\Pi_\theta(k) \sim (1+ k/k_c) \exp(-k/k_c).
\label{eq:Eflux_exponential}
\end{equation}
Our above arguments closely resemble the derivation of the energy flux for small-Re flows~\cite{Verma:FD2018}, and for quasi-static magnetohydrodynamic turbulence with strong interaction parameters~\cite{Verma:PF2015QSMHD}.

Figure~\ref{fig:Entropy_smallPr}(a) shows that the lower branch of the entropy spectrum fits well with Eq.~(\ref{eq:Entropy_smallPr_proposal}) in the intermediate wavenumbers, with $k_c=32$ for $\mathrm{Pr}=0.1$ and $k_c=25$ for $\mathrm{Pr}=0.02$. Further, as evident from Fig.~\ref{fig:Entropy_smallPr}(b), the entropy fluxes for $\mathrm{Pr}=0.1$ and $0.02$ obey Eq.~(\ref{eq:Eflux_exponential}). Our results are consistent with earlier studies \cite{Mishra:PRE2010,Verma:book:BDF,Verma:book:ET} that also obtained similar exponential scalings in the entropy spectrum and flux of small-Pr convection.

%

\end{document}